\documentclass[sigconf]{acmart}
\usepackage{accsupp}

\AtBeginDocument{%
  }

\copyrightyear{2026}
\acmYear{2026}
\setcopyright{cc}
\setcctype{by-nc-nd}
\acmConference[CHI EA '26]{Extended Abstracts of the 2026 CHI Conference on Human Factors in Computing Systems}{April 13--17, 2026}{Barcelona, Spain}
\acmBooktitle{Extended Abstracts of the 2026 CHI Conference on Human Factors in Computing Systems (CHI EA '26), April 13--17, 2026, Barcelona, Spain}
\acmDOI{10.1145/3772363.3798818}
\acmISBN{979-8-4007-2281-3/2026/04}

\begin{document}

\title{TactileWalk: Dynamic Electrotactile Patterns for Fingertip-Based Interaction During Walking}
\author{Vedika Nimbalkar}
\email{vn4032@rit.edu}
\affiliation{%
  \institution{AIR Lab, School of Information, GCCIS, \\Rochester Institute of Technology}
   \city{Rochester}
   \state{NY}
   \country{USA}
}

\author{Roshan Peiris}
\email{roshan.peiris@rit.edu}
\affiliation{%
  \institution{AIR Lab, School of Information, GCCIS, \\Rochester Institute of Technology}
   \city{Rochester}
   \state{NY}
   \country{USA}
}

\renewcommand{\shortauthors}{Nimbalkar and Peiris}

\begin{abstract}

TactileWalk evaluates dynamic electrotactile patterns on fingertips for wearable navigation. We developed a fingertip stimulation prototype featuring a $10 \times 6$ electrode grid driven by an ESP32 microcontroller and high-voltage drivers to enable rapid, independent electrode activation for spatiotemporal pattern rendering. This research compares three dynamic patterns: Single Line, Double Line, and Box—across eight directions presented on the tactile display at the fingertip. Study 1 (stationary) revealed that simple linear patterns were recognized significantly more accurately than complex shapes. Study 2 (walking) confirmed these cues remain robust under movement, where the Double Line pattern yielded the highest accuracy (90.83\%). Participants consistently preferred the "reinforcing" Double Line and found vertical motion more intuitive while walking. We propose design implications for mobile haptics, advocating for simple, spatially redundant patterns to minimize cognitive load during eyes-free navigation.

\end{abstract}




\maketitle

\section{Introduction}
Haptic feedback is a key component of human–computer interaction, enabling users to perceive information through physical sensations rather than visual or auditory cues. It has become increasingly important for immersive interaction in emerging systems such as augmented reality (AR), virtual reality (VR), and wearable devices.

\begin{figure}[t]
    \centering
    \includegraphics[width=.8\linewidth]{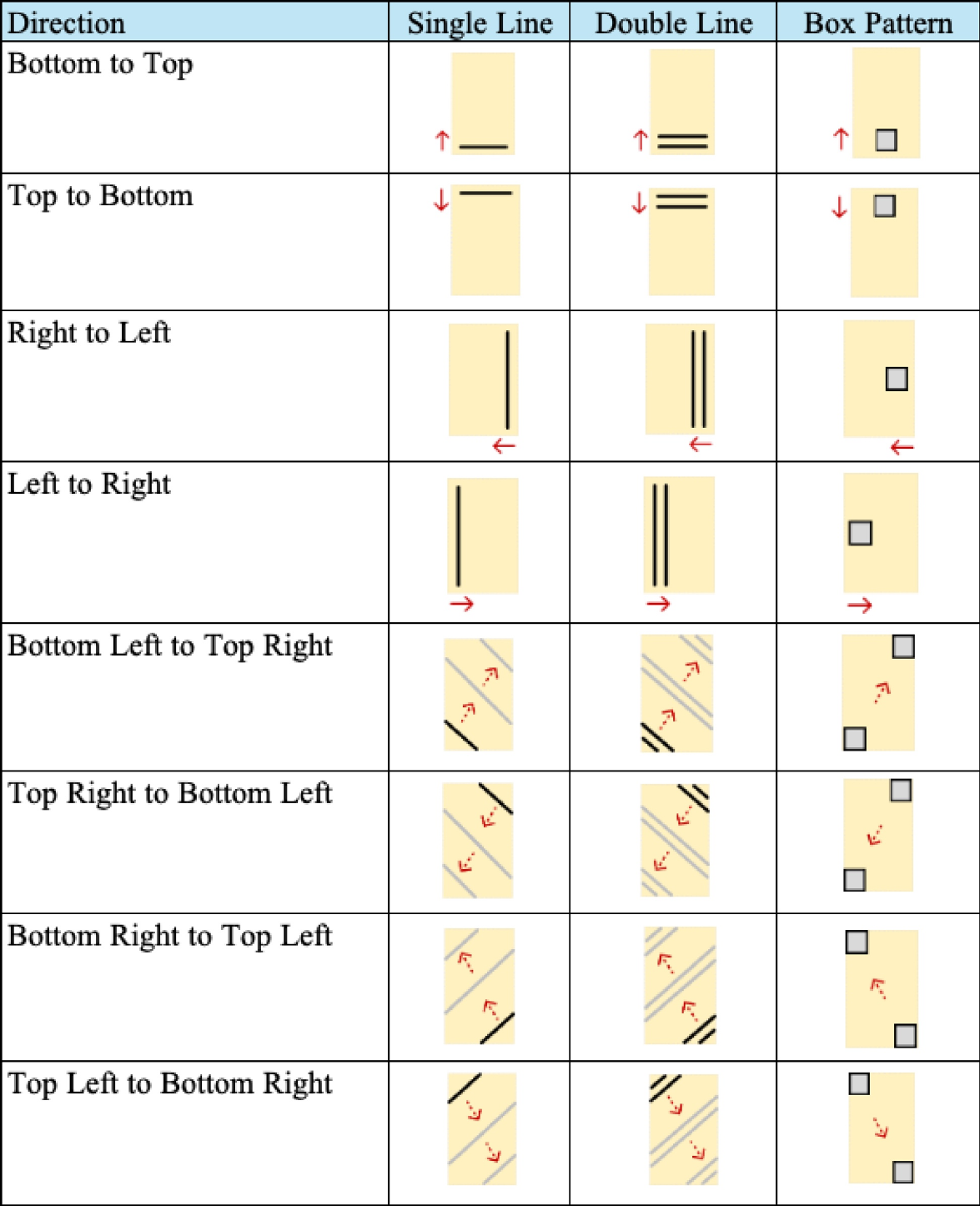}
     \Description{Table illustrating the electrotactile pattern set, showing three pattern types (Single Line, Double Line, and Box) across eight directions on a fingertip electrode grid.}
    \caption{Electrotactile pattern taxonomy for Study 1. The stimulus set includes three fundamental spatial primitives: (A) Single Line, (B) Double Line, and (C) Box patterns. Arrows indicate the dynamic movement patterns applied to each configuration during the study.}
    \label{fig:patterns}
\end{figure}

Electrotactile feedback, in particular, has received growing attention due to its potential for precision and practicality in wearable and mobile applications \citep{vizcay2021electrotactile}. Electrotactile stimulation is the process of generating tactile sensations by applying electrical currents to stimulate nerves \citep{palmarside}. This technique works by passing an electric current through the skin, creating localized sensations at the site of the electrodes \citep{68204}. Research has demonstrated the versatility of electrotactile stimulation in applications such as AR and VR \cite{vizcay2021electrotactile, 10.1016/j.cag.2023.01.013, Encodingtactile, selfpowered, forcesensing, shapeelectro}, skill training ~\citep{8906989, 10.1007/978-3-030-58147-3_49}, textile rendering \cite{TacTex}, and assistive technologies ~\citep{10.1145/3613904.3642546, LUO2023108425, Shehata2018-zk, swaybased, graphElectro}.

Electrotactile feedback has been investigated across diverse body locations—including the feet \citep{FeetThrough}, lips \citep{LipIO}, tongue \citep{TactTongue}, arms \citep{iFeel}, and wrists \citep{ShockMeTheWay}—to support applications in navigation, accessibility, texture rendering \citep{TacTex}, material perception \citep{Material}, and virtual reality \citep{vizcay2021electrotactile, selfpowered}. Much of this research has utilized sparse or linear electrode arrays to explore perceptual phenomena such as apparent motion, slip illusions, and funneling effects on the forearm and fingertips \citep{okabe2011fingertip, sparks1979identification, NEEDLE, funnel}.

The fingertip is a particularly compelling site for high-resolution feedback due to its high mechanoreceptor density. While prior work has established that vertical fingertip movements are more easily recognized than horizontal or diagonal patterns \citep{Spatiotemporal}, there remains significant potential for using high-density arrays to convey complex spatiotemporal information for applications such as navigation.

In the context of mobile navigation, providing feedback through alternative sensory channels can reduce cognitive load by minimizing reliance on visual or auditory cues \citep{ShockMeTheWay}. While existing navigation research has explored vibrotactile \citep{hapticCollar, syncRunning, vibNav}, thermal \citep{thermalBracelet, thermalCue}, and electrotactile modalities at the wrist \citep{ShockMeTheWay} and feet \citep{FeetThrough, MetaAnalysis}, the fingertip offers a unique advantage for providing electrotactile feedback for navigation: It enables the delivery of precise, dynamic navigational cues while the user is already naturally interacting with a mobile device during navigation.

To explore this opportunity, we conducted two studies to evaluate the perceptibility of different spatiotemporal patterns and to assess their suitability for navigation. Hence, in Study 1, we designed three types of dynamic electrotactile patterns: Single Line, Double Line, and Box, each moving in eight directions (Figure~\ref{fig:patterns}), and evaluated their perceptual accuracy in a stationary setting, addressing \textit{RQ1: How accurately can electrotactile patterns be perceived on the fingertips?}. Building on the findings, Study 2 assessed whether these patterns remain interpretable while walking, addressing \textit{RQ2: How does walking influence the accuracy and response time of electrotactile pattern recognition?} Together, these studies evaluate the suitability of dynamic fingertip-based electrotactile patterns for navigation-oriented wearable and mobile interaction.

\section{Study 1: Perception of Dynamic Electrotactile Patterns on the Fingertip}

\begin{figure}[htpb]
    \centering
    \includegraphics[width=.85\linewidth]{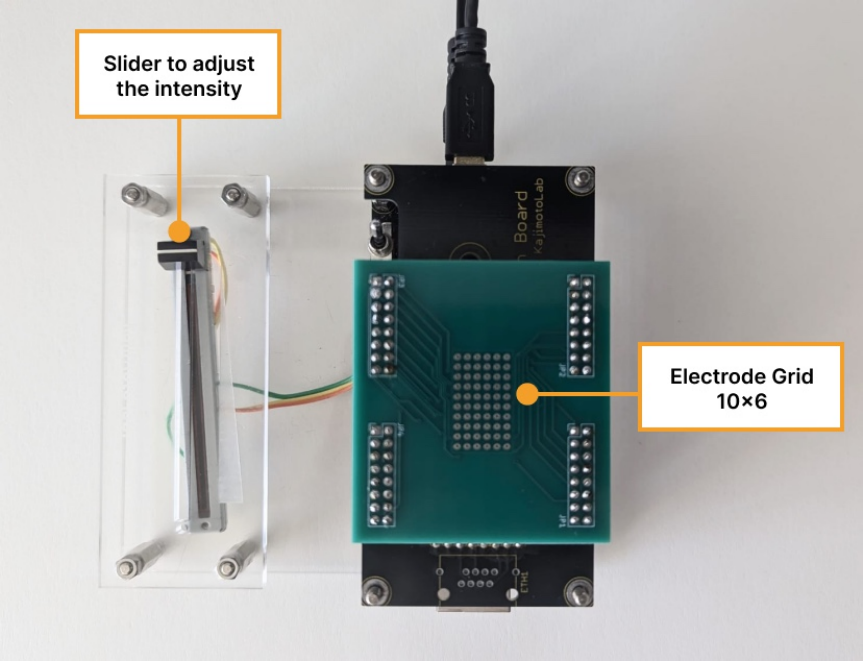}
    \fbox{\includegraphics[width=.8\linewidth]{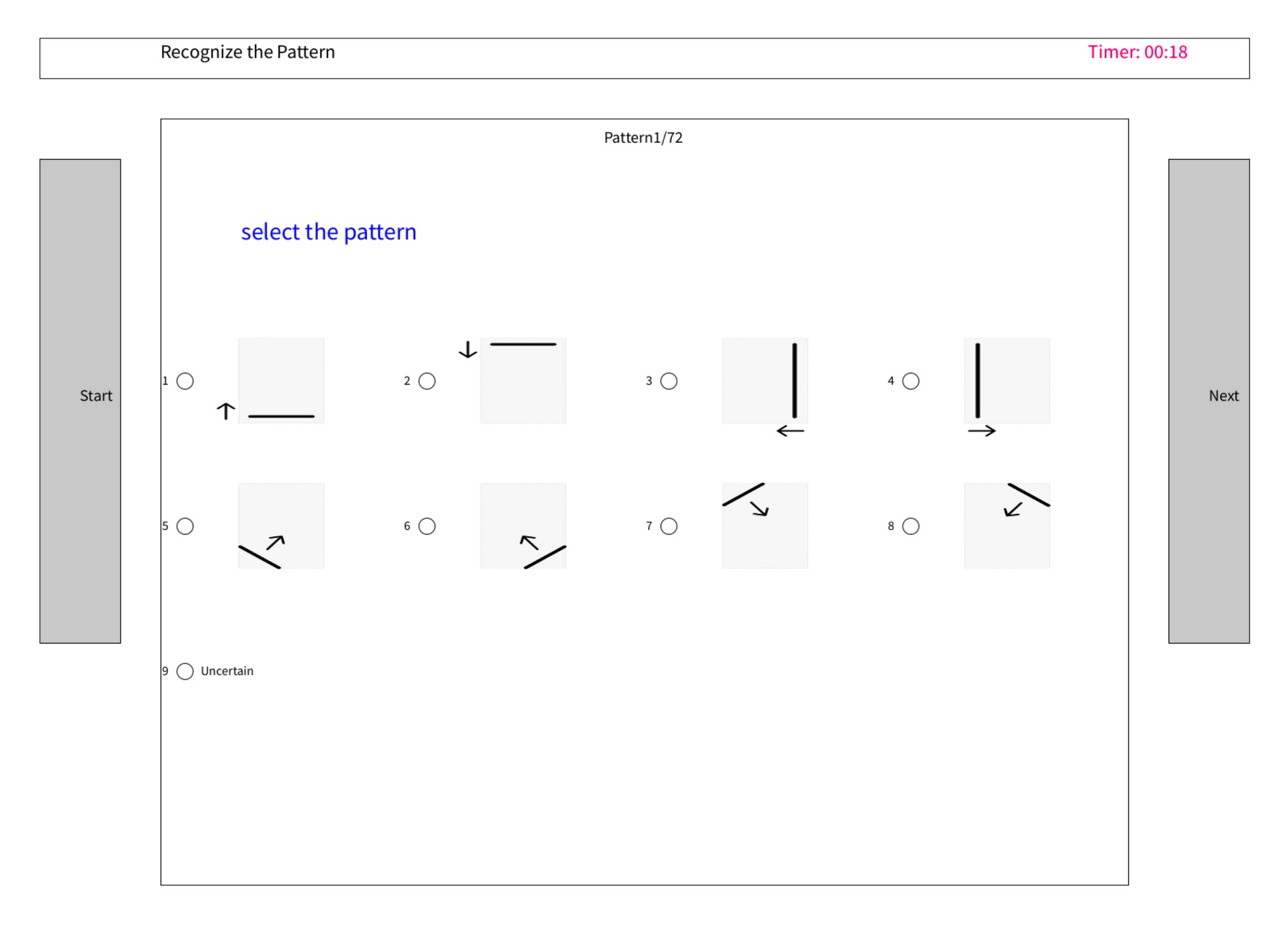}}
    \includegraphics[width=.85\linewidth]{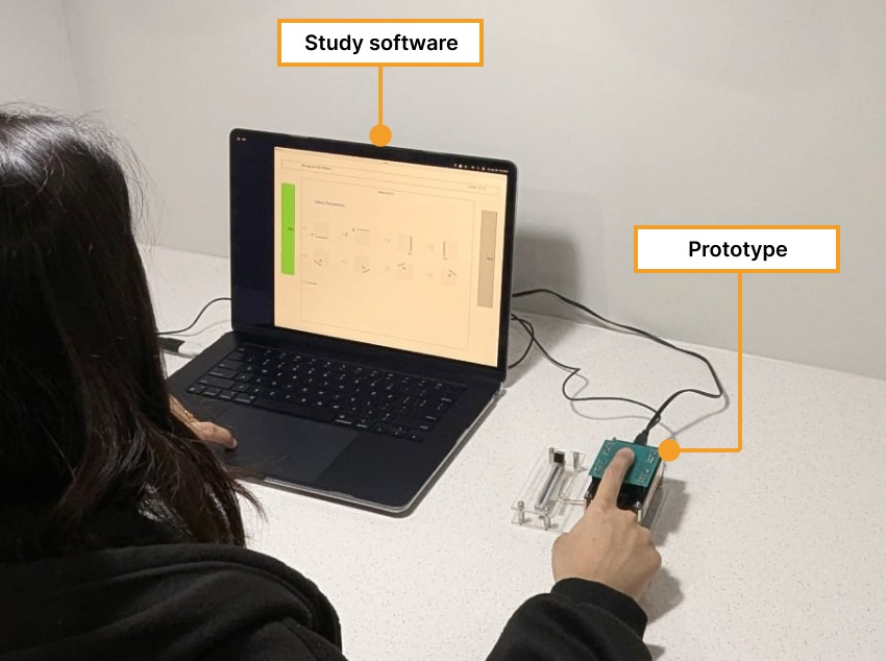}
    \Description{Overview of Study 1 materials. Top: Table illustrating the electrotactile pattern set, showing three pattern types (Single Line, Double Line, and Box) across eight directions on a fingertip electrode grid. Center: Software used by participants for the evaluation. It displays a Start button, the 8 movements of the haptic patterns with radio buttons next to them and a Next button. Bottom: The image shows a participant seated in front a laptop that displays the study software and has their fingertip on the electrode pad.}
    \caption{(Top) The fingertip electrotactile prototype set up. This interface consists of a 10×6 electrode grid and intensity control slider. (Center) software interface used to present stimuli and collect participant responses. Clicking `Start' button starts the stimuli on the fingertip and the participant is required to select a perceived pattern and click `Next' to confirm. (Bottom) Study setup where a participant experiences the feedback on the fingertip and reports their results}
    \label{fig:study1_prototype}
\end{figure}

\textit{Prototype: }Figure \ref{fig:study1_prototype} illustrates the electrotactile feedback system, which utilized a fingertip stimulation device adapted from prior work by \citet{PWMPrototype, Kit}. We customized the prototype to feature a $10 \times 6$ electrode grid consisting of circular electrodes of $1$ mm diameter with $1$ mm inter-electrode spacing. The grid was driven by a multi-channel system controlled via an ESP32 microcontroller and high-voltage shift-register drivers (HV513/HV507), enabling rapid, independent electrode addressing. The system operated at approximately $300$ V using a current-regulated architecture.

The stimulation was configured as monopolar: at each stimulation frame, a single electrode was activated as the cathode (active), while all remaining electrodes in the grid were connected to the complementary reference line acting as a collective anode (return). To render spatial patterns, the system employed a sequential scanning method rather than simultaneous activation; each electrode in the intended pattern was assigned its role momentarily before moving to the next. This spatiotemporal rendering utilized monophasic, current-controlled pulses with a fixed pulse width of 100 $\mu$s.

Control software included within the study software (Fig~\ref{fig:study1_prototype}(center))-implemented in Processing-delivered patterns with a consistent 25 ms inter-stimulus interval (ISI). This value was selected based on pilot testing, which determined that 25 ms effectively balanced the requirements for apparent motion—ensuring the sensation felt like a continuous moving line rather than discrete points-while avoiding the "temporal blending" or discomfort associated with faster refresh rates or hardware-induced latency. Stimulation amplitude was individually calibrated for each participant to ensure a clearly perceivable yet comfortable level, with current regulation maintaining consistency despite variations in skin impedance. 

\textit{Pattern Design: }Using this system, we evaluated three dynamic electrotactile patterns: Single Line, Double Line, and Box, each presented in eight movement directions: Bottom to Up, Up to Down, Left to Right, Right to Left, Bottom Left to Top Right, Top Right to Bottom Left, Bottom Right to Top Left, and Top Left to Bottom Right. Pattern selection was informed by prior work~\citep{Spatiotemporal} and pilot testing to balance simplicity and expressiveness: Single Line for minimal cues, Double Line to examine spatial redundancy, and Box to assess perception of a moving solid shape. Pattern type served as the primary independent variable, with each direction repeated three times, yielding 72 trials per participant; pattern types and direction order were fully randomized.

The number of activated electrodes per stimulation frame depended on pattern type and movement direction. For the Single Line pattern, activation corresponded to one full row or column of the 10×6 array, with horizontal movements activating 10 electrodes per frame and vertical movements activating 6. For diagonal movements, the number of active electrodes varied dynamically as the line traversed the grid, increasing from a single electrode at the corner to a maximum span and decreasing symmetrically as it exited. The Double Line pattern activated two immediately adjacent rows or columns, resulting in exactly twice the number of active electrodes as the corresponding Single Line condition. The Box pattern activated a 4×3 rectangular contour (n = 12 electrodes per frame), which remained constant across movement directions.

\textit{Procedure: } Twelve participants (aged 22–28 years, M = 24.92, SD = 1.73) were recruited from the institution and used their preferred hand to interact with the prototype. Participants primarily placed the index finger of their preferred hand on the electrode grid. After providing informed consent and receiving a brief introduction, participants completed a short familiarization phase to select a comfortable stimulation intensity. Stimulation amplitude was individually calibrated for each participant to a clearly perceivable yet comfortable level prior to the experimental trials. Each participant then completed 72 randomized trials (three repetitions of each pattern across eight directions), with breaks provided as needed to reduce fatigue. Following each pattern type, participants reported confidence levels, completed a NASA TLX questionnaire~\citep{NASATlx}, and answered brief experience-related questions. A post-study interview was conducted to collect qualitative feedback on usability and overall experience.

\subsection{Results of Study 1}
\begin{figure}[h]
    \centering
    \includegraphics[width=0.95\linewidth]{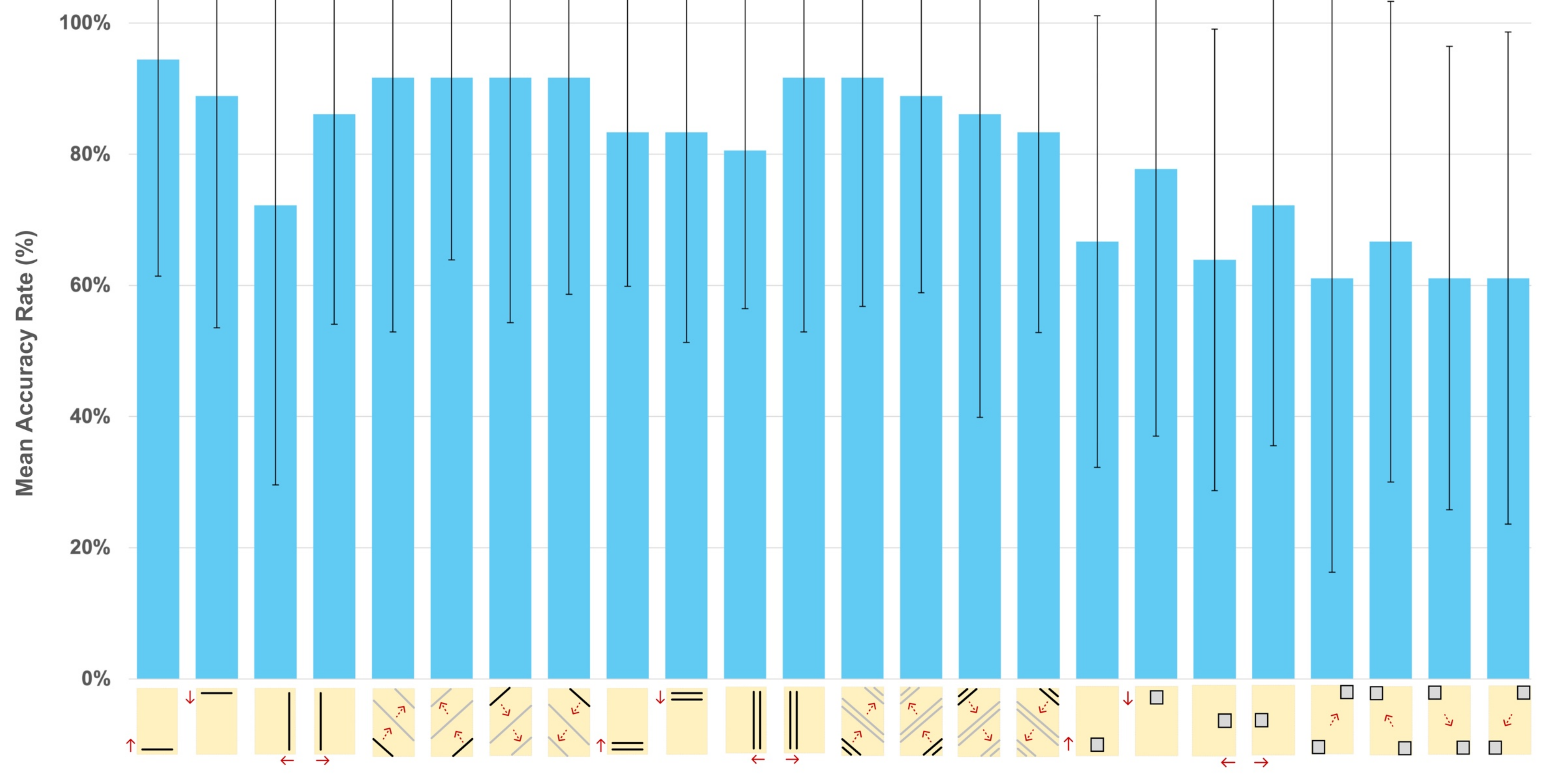}
    \hfill
    \includegraphics[width=0.95\linewidth]{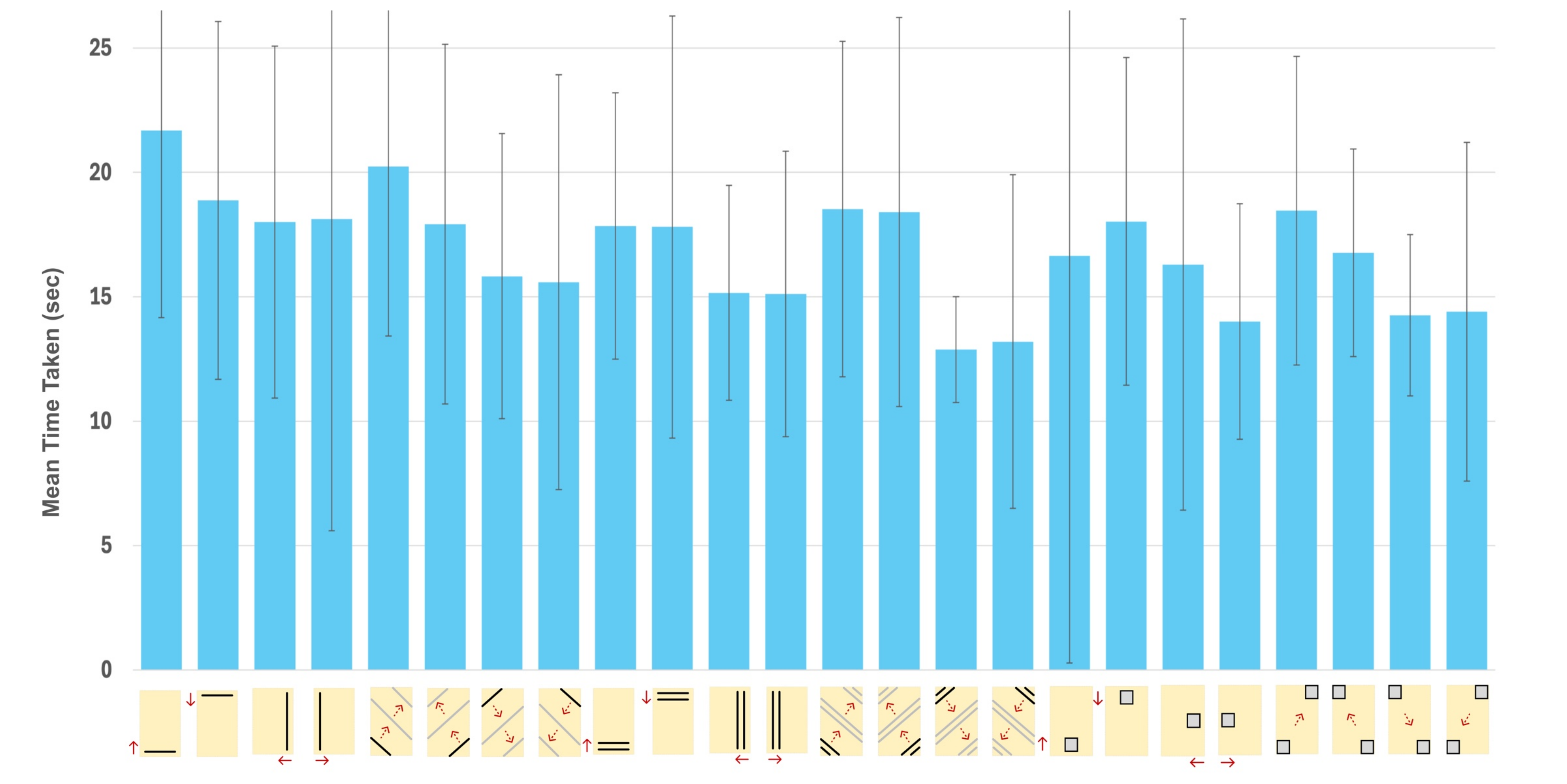}
    \Description{Bar chart showing Study 1 results. Top: Mean accuracy rates for Single Line, Double Line and Box patterns across eight directions. Single Line and Double Line show higher accuracy than Box patterns. Bottom: Mean response times for the same patterns and directions, showing comparable response times across pattern types with greater variability for the Box pattern.}
    \caption{Study 1 results showing mean accuracy of the perceived pattern (Top) and mean response time (Bottom). Error bars represent standard deviation.}
    \label{fig:study1_results}
\end{figure}

Figure~\ref{fig:study1_results} shows the mean perceived accuracy for all pattern types. 
Overall accuracy was highest for the Single Line pattern (88.54\%), followed by the Double Line pattern (86.11\%), while the Box pattern showed substantially lower accuracy (66.32\%). A one-way repeated-measures ANOVA was conducted with Pattern Type (Single Line, Double Line, Box) as the within-subject factor. Mauchly’s Test of Sphericity was violated ($\chi^2(2) = 7.147, p < 0.05$). Therefore, Greenhouse-Geisser corrected repeated-measures ANOVA revealed a significant main effect of pattern type on accuracy $(F_{(1.35,14.88)} = 8.45, p = 0.007)$. Post-hoc Bonferroni comparisons indicated that both Single Line and Double Line patterns were recognized significantly more accurately than the Box pattern ($p < .003$ and $p < .009$, respectively), with no significant difference between Single Line and Double Line ($p = 1.00$).

The mean time taken for all directions are shown in the Figure~\ref{fig:study1_results}. 
Although the Single Line pattern had the highest accuracy, it also took the longest overall to complete, with an average time of 1828.25 ms. In contrast, the Double Line and Box Patterns had similar completion times, averaging 1611.56 ms and 1610.73 ms, respectively. For the time taken for each pattern type, a Repeated Measures ANOVA test was conducted after verifying that the assumptions of normality (Shapiro-Wilk tests, $p > .05$) and sphericity (Mauchly’s test, $p > .05$) were satisfied. The results showed no significant effect of pattern type on completion time, $F_{(2,22)}=3.422, p = 0.051$.  

Participant feedback largely aligned with the quantitative findings. While the Box pattern yielded the lowest accuracy, participants expressed mixed reactions: 3 out of 12 participants found its compact stimulation area helpful, whereas others reported difficulty in discerning movement direction, particularly for diagonal patterns (e.g., \textit{``the directions for box pattern are kind of difficult to tell''} (P14)). In contrast, participants consistently reported positive experiences with the Single Line and Double Line patterns. 2 out of 12 participants noted that the Double Line pattern provided clearer directional confirmation, with one stating that \textit{``the second line would act like the confirmation for the first movement''} (P5). Participants also reported greater difficulty recognizing diagonal movements than vertical or horizontal ones, especially for the Box pattern, suggesting that complex shapes combined with diagonal motion reduce perceptual clarity.
\section{Study 2: Perception of Electrotactile Patterns on the Fingertip While Walking}

\begin{figure}[h]
    \centering
    \raisebox{-0.5\height}{    \includegraphics[width=0.4\linewidth]{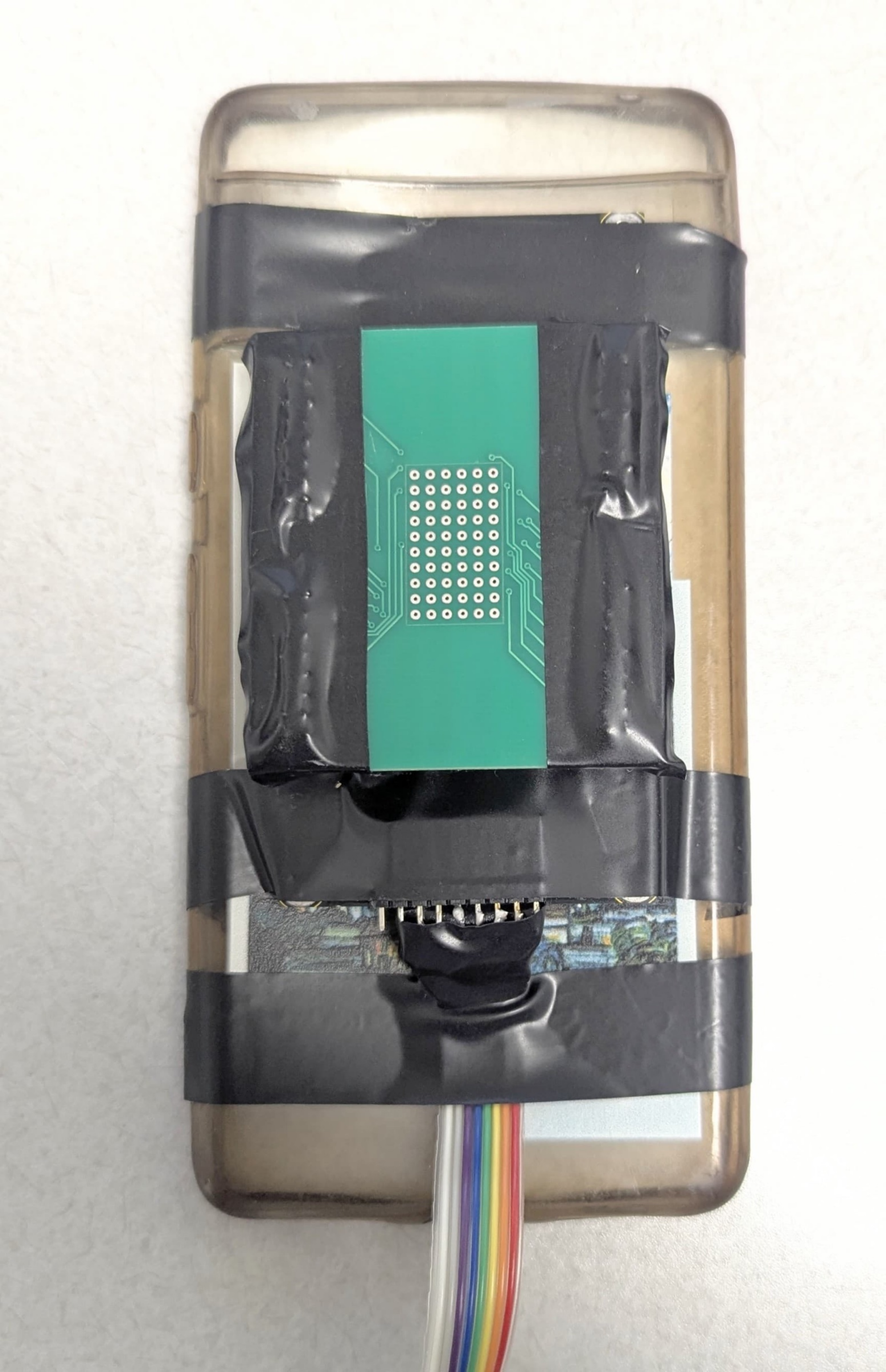}}
    \raisebox{-0.5\height}{ 
    \includegraphics[width=0.5\linewidth]{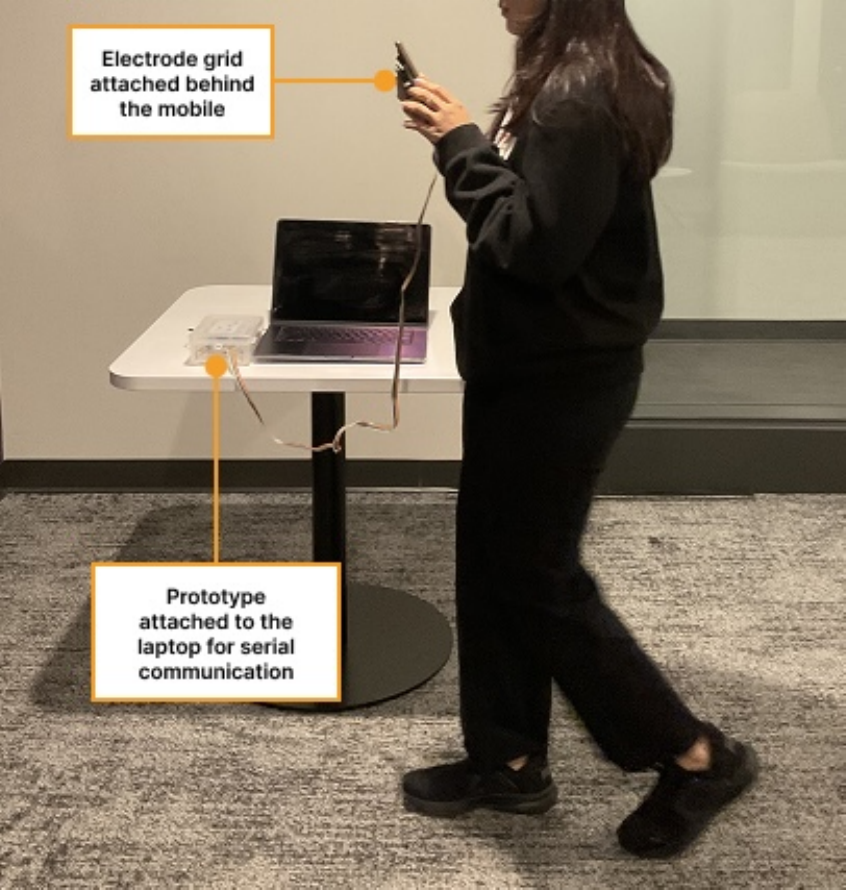}}
   \Description{Two images illustrating the Study 2 setup. Left: A mobile phone with a fingertip electrotactile electrode grid mounted on the back. Right: A participant walking while holding the mobile phone and performing the pattern recognition task, with the electrotactile prototype connected to the laptop via a wired cable for control and data logging.}
    \caption{Setup of the 2nd study with the mobile setup. (Left) The electrotactile grid mounted on the back of the phone. (Right) Study setup showing a participant performing the walking-based pattern recognition task. Participant's mobile device displayed an interface similar to Fig.~\ref{fig:study1_prototype}(center) to report their responses.}
    \label{fig:study2_prototype}
\end{figure}

Building on the results of Study 1, Study 2 evaluated the robustness of electrotactile pattern recognition during walking by examining changes in accuracy and response time under movement.

\textit{Prototype and Pattern Design.} Study 2 used the same electrotactile hardware as Study 1, with the 10×6 electrode grid mounted on the back of a mobile phone to stimulate the fingertip (Figure~\ref{fig:study2_prototype}). A Processing-based Android application controlled stimulus presentation and response input, communicating with the electrotactile controller via a wired serial connection and automatically logging responses and response times to a CSV file. All patterns were presented with a fixed duration and a consistent 25 ms inter-step delay. Building on Study 1, pattern type remained the primary independent variable but was reduced to two conditions: Single Line and Double Line. Based on pilot testing, diagonal directions were excluded due to poor recognizability during walking, resulting in four directions (Bottom to Top, Top to Bottom, Left to Right, Right to Left). Each direction was repeated three times, yielding 24 randomized trials per participant.


\textit{Procedure.} The study involved 10 participants (aged 24–29 years, M = 25.1, SD = 1.52), recruited from the institution. Participants used their preferred hand and walked at a comfortable, self-selected pace to support natural interaction. After providing consent and receiving a brief refresher, participants performed the pattern recognition task while walking back and forth along a 3–4 ft straight path, necessitated by the limited length of the wired connection between the electrotactile pad and the laptop. The electrotactile grid was mounted on the back of a mobile phone, which participants held naturally while keeping their index finger on the pad; the second hand could be used for stabilization as needed (Figure~\ref{fig:study2_prototype}). After completing trials for each pattern type, participants reported confidence levels, completed a NASA TLX questionnaire~\citep{NASATlx}, and answered brief experience-related questions. A post-study interview captured qualitative feedback on usability.

\subsection{Results of Study 2}
\begin{figure}[h]
    \centering
    \includegraphics[width=0.9\linewidth]{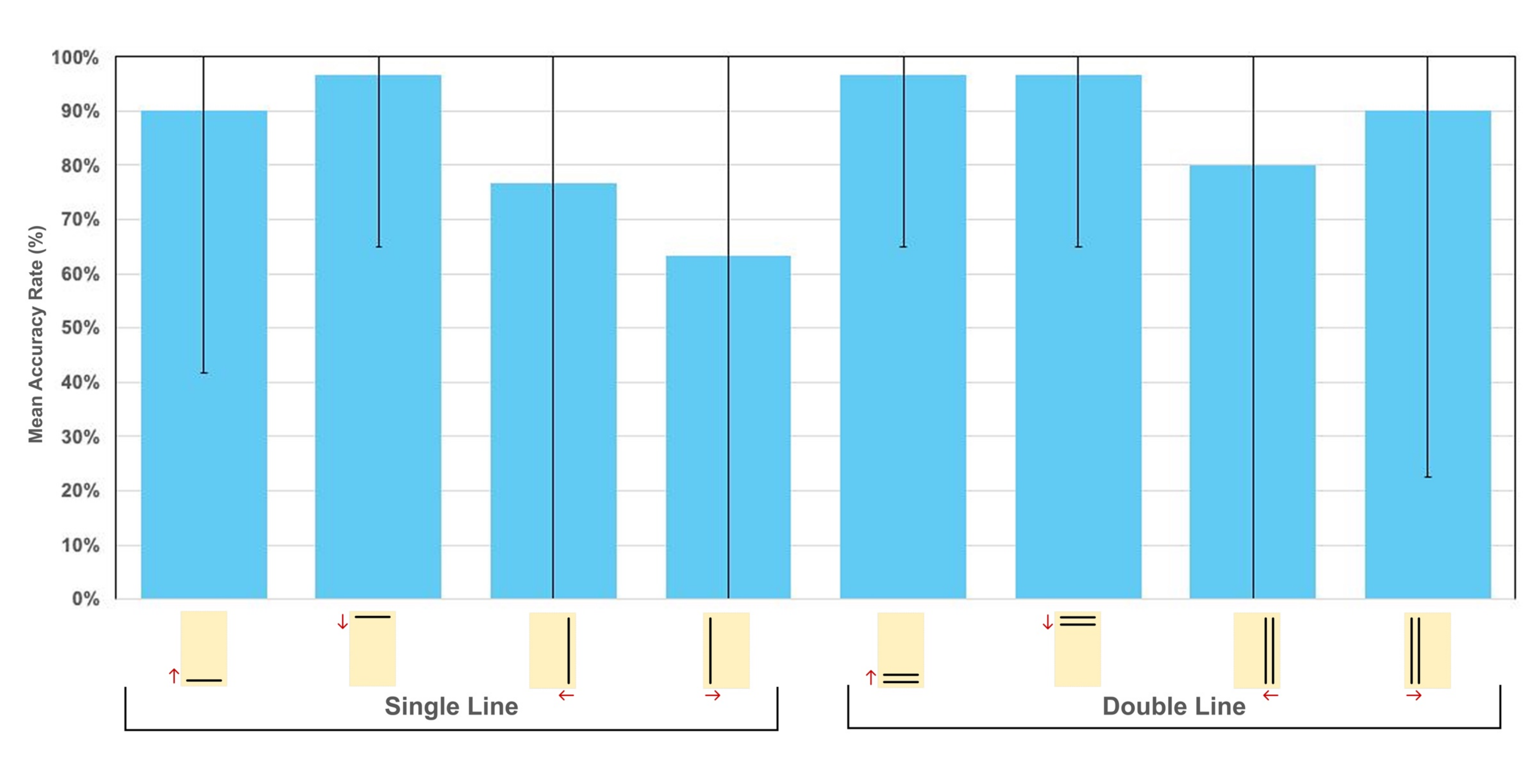}
    \includegraphics[width=0.9\linewidth]{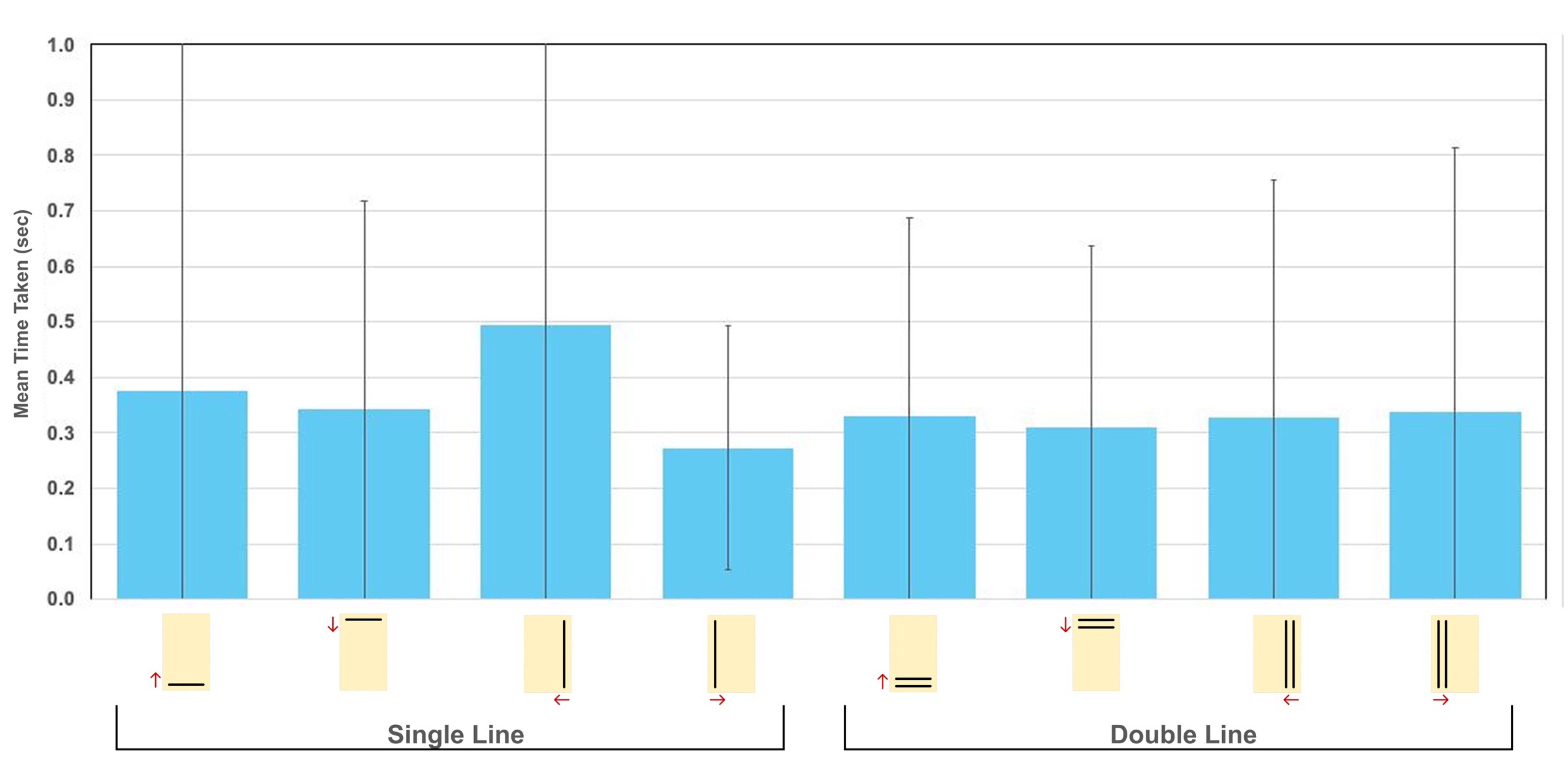}
    \Description{Two bar charts showing Study 2 results. Top: Mean accuracy rates for Single Line and Double Line across four directions. Double Line showing consistently higher accuracy and lower variability. Bottom: Mean response times for the same patterns and directions, indicating similar response times across patterns with reduced variability for the Double Line pattern. Error bars indicate standard deviation.}
    \caption{Study 2 results showing mean accuracy (left) and response time (right) during walking. Error bars represent standard deviation.}
    \label{fig:study2_results}   
\end{figure}

Figure~\ref{fig:study2_results} shows the mean accuracy for the two pattern types while walking. Participants achieved a higher average accuracy with the Double Line pattern ($90.83\%$) compared to the Single Line pattern ($81.67\%$). The Double Line pattern consistently performed better across all directions, with 60\% of participants achieving perfect accuracy. Due to the small sample size ($N=10$) and negatively skewed data in the Double Line condition, a Wilcoxon Signed-Rank test was employed. The analysis revealed no statistically significant difference between the two conditions ($Z = 8, p = 0.182$), though the observed trend aligns with participant preferences for the reinforced directional cues of the Double Line pattern.


Mean response times are shown in Figure~\ref{fig:study2_results}. The Single Line pattern had an average response time of 351 ms, while the Double Line pattern showed a slightly faster average of 326 ms. Response times were generally stable across directions, although the Single Line pattern exhibited greater variability. A Wilcoxon Signed-Rank test was utilized for the analysis after a Shapiro-Wilk test confirmed that the data was not normally distributed ($p < 0.001$). The analysis revealed no statistically significant difference between the two conditions ($Z = 22, p = 0.625$) These results suggest that both patterns were recognized with comparable speed during walking, despite the slightly faster responses observed for the Double Line pattern.

Overall, participants provided positive feedback on the use of the electrotactile system while walking. 7 out of 10 participants reported that they would consider using such a system for everyday navigation, and one participant suggested a wearable implementation, stating that they \textit{``would also use it behind the smartwatch''} (P6). Participants consistently expressed a preference for the Double Line pattern, describing it as clearer and easier to interpret than the Single Line; one participant noted that \textit{``with both the lines giving stimulation, it gave me a better idea of which direction it was going''} (P2). This preference was reflected in performance outcomes, with 6 out of 10 participants achieving perfect accuracy in the Double Line condition. In terms of direction, participants generally found vertical movements (Top to Bottom and Bottom to Top) more intuitive than horizontal movements, which were often described as less distinct (e.g., \textit{``horizontal directions are a bit confusing to recognize''} (P4)). This feedback aligns with prior findings that vertical electrotactile motion on the fingertip is perceived more reliably than horizontal motion \citep{Spatiotemporal}.

\section{Discussion}
Across two studies, we investigated how dynamic electrotactile patterns on the fingertip are perceived under stationary and walking conditions. Study 1 showed that simple linear patterns, Single Line and Double Line, were recognized significantly more accurately than the Box pattern, indicating that pattern simplicity and spatial continuity are critical for fingertip-based electrotactile perception. Consistent with prior work \cite{Spatiotemporal}, diagonal movements were harder to recognize than vertical and horizontal ones, highlighting the importance of aligning pattern motion with the fingertip’s sensitivity profile.

Building on these findings, Study 2 examined pattern recognition during walking and found that electrotactile cues remained interpretable despite movement. Simplifying the pattern set to two patterns and four directions reduced cognitive load and supported reliable recognition. The Double Line pattern again performed strongly, yielding higher accuracy, greater confidence, and stronger user preference, with participants describing the second line as a reinforcing directional cue. Vertical movements were perceived as more intuitive than horizontal ones, extending earlier findings \cite{Spatiotemporal} to mobile contexts.

Together, these results align with prior work demonstrating the robustness of electrotactile feedback during movement, such as wrist-based navigation while cycling \cite{ShockMeTheWay}, and further support the fingertip as an effective site for high-resolution electrotactile feedback \cite{Tacttoo}. By evaluating dynamic spatiotemporal patterns across stationary and walking conditions, this work extends existing research and highlights the potential of fingertip-based electrotactile feedback for wearable and navigation applications. Based on findings from both studies, we conclude that electrotactile navigation systems for mobile contexts should prioritize simple linear patterns, such as the Single and Double Line, over complex shapes like the Box pattern to ensure higher recognition accuracy. Designers should leverage reinforced patterns like the Double Line to improve perceptual clarity and user confidence during movement. Furthermore, pattern and direction sets should be minimized to reduce cognitive load and maintain robust interpretation while walking.



\subsection{Limitations and Future Work}
This work was conducted with a relatively small sample size in controlled indoor environments, which may limit generalizability to more complex real-world settings. Additionally, while pattern simplification improved recognition during walking, future studies should further explore how diagonal and more complex patterns can be made more perceivable. 

A key advantage of the proposed approach is that it provides non-visual, non-auditory guidance, which can reduce reliance on sensory channels that are often overloaded during mobile navigation. Although a walking condition was included, the task primarily focused on pattern recognition rather than realistic navigation. Real-world navigation typically involves additional cognitive demands—such as obstacle avoidance, route planning, and environmental distractions—which may affect users’ ability to perceive and interpret haptic cues. Consequently, the present findings may overestimate performance in everyday settings. Future work should evaluate the technique in ecologically valid wayfinding scenarios (e.g., indoor or outdoor routes with competing attentional demands) to better assess its effectiveness under real-world conditions.

\section{Conclusion}

This work examined how dynamic electrotactile patterns on the fingertip are perceived under both stationary and walking conditions. Across two studies, simple linear patterns consistently outperformed more complex shapes, with the Double Line pattern providing the highest accuracy, faster responses, and greater user confidence. Importantly, electrotactile cues remained interpretable during walking, demonstrating robustness under movement. These findings highlight the fingertip as an effective site for directional electrotactile feedback and support the use of simple, reinforced patterns for wearable and mobile navigation applications.




\bibliographystyle{ACM-Reference-Format}
\bibliography{references}


\end{document}